%% file: main.tex
\documentclass{llncs}

\usepackage{times}
\usepackage{latexsym}
\usepackage{paralist}
\usepackage{citesort}
\usepackage{amssymb}

\input{macros}

\begin{document}

\title{Higher-Order Termination:\\ From Kruskal to
Computability}

\author{
  Fr{\'e}d{\'e}ric Blanqui\inst{1}
  \and
  Jean-Pierre Jouannaud\inst{2}\thanks{Project
  LogiCal, P{\^o}le Commun de Recherche en Informatique du Plateau de
  Saclay, CNRS, {\'E}cole Polytechnique, INRIA, Universit{\'e} Paris-Sud.}
  \and
  Albert Rubio\inst{3}}
 
\institute{
  {INRIA \& LORIA, BP 101, 54602 Villiers-l{\'e}s-Nancy CEDEX, France}
\and
  {LIX, {\'E}cole Polytechnique,
  91400 Palaiseau, France}
\and
  {Technical University of Catalonia,
  Pau Gargallo 5, 08028 Barcelona, Spain}  
}

\maketitle

\section{Introduction}
\label{s:introduction}

Termination is a major question in both logic and computer science. In
logic, termination is at the heart of proof theory where it is usually
called strong normalization (of cut elimination). In computer science,
termination has always been an important issue for showing programs
correct. In the early days of logic, strong normalization was usually
shown by assigning ordinals to expressions in such a way that
eliminating a cut would yield an expression with a smaller ordinal. In
the early days of verification, computer scientists used similar
ideas, interpreting the arguments of a program call by a natural
number, such as their size.  Showing the size of the arguments to
decrease for each recursive call gives a termination proof of the
program, which is however rather weak since it can only yield quite
small ordinals. In the sixties, Tait invented a new method for showing
cut elimination of natural deduction, based on a \emph{predicate} over
the set of terms, such that the membership of an expression to the
predicate implied the strong normalization property for that
expression. The predicate being defined by induction on types, or even
as a fixpoint, this method could yield much larger ordinals.  Later
generalized by Girard under the name of \emph{reducibility} or
\emph{computability candidates}, it showed very effective in proving
the strong normalization property of typed lambda-calculi with
polymorphic types, dependent types, inductive types, and finally a
cumulative hierarchy of universes.  On the programming side, research
on termination shifted from programming to executable specification
languages based on rewriting, and concentrated on automatable methods
based on the construction on well-founded orderings of the set of
terms. The milestone here is Dershowitz's \emph{recursive path
ordering} (RPO), in the late seventies, whose well-foundedness proof
is based on a powerful combinatorial argument, Kruskal's tree
theorem, which also yields rather large ordinals. While the
computability predicates must be defined for each particular case, and
their properties proved by hand, the recursive path ordering can be
effectively automated.

These two methods are completely different. Computability arguments show
\emph{termination}, that is, infinite decreasing sequences of
expressions $e_0\succ e_1\succ\ldots e_n\succ e_{n+1}\ldots$ do not
exist. Kruskal's based arguments show \emph{well-orderedness}: for any
infinite sequence of expressions $\{e_i\}_i$, there is a pair $j<k$
such that $e_j\preceq e_k$. It is easy to see that well-orderedness
implies termination, but the converse is not true.

In the late eighties, a new question arose: termination of a
simply-typed lambda-calculus language in which beta-reduction would be
supplemented with terminating first-order rewrite rules. Breazu-Tannen
and Gallier on the one hand~\cite{BTG90}, and
Okada~\cite{okada89issac} on the other hand, showed that termination
was satisfied by the combination
by using computability arguments. Indeed, when rewriting
operates at basic types and is generated by first-order rewrite
rules, beta-reduction and rewriting do not interfere. Their result,
proved for a polymorphic $\lambda$-calculus, was later generalized to
the calculus of constructions~\cite{barbanera90}.  The situation becomes
radically different with higher-order rewriting generated by rules
operating on arrow-types, or involving lambda-bindings or higher-order
variables. Such an example is provided by G{\"o}del's system
$T$, in which higher-order primitive recursion for natural
numbers generated by Peano's constructors $0$ and $s$ is described by
the following two higher-order rules :
\begin{eqnarray*}
rec(0,U,V) &\ra& U\\
rec(s(X),U,V) &\ra& @(V,X,rec(X,U,V))
\end{eqnarray*}
where $rec$ is a function symbol of type $\Nat \ra
T \ra (\Nat\ra T \ra T)\ra T$, $U$ is a higher-order variable of type
$T$ and $V$ a higher-order variable of type $\Nat \ra T \ra T$, for all type $T$.
Jouannaud and Okada invented the so-called
general-schema~\cite{jouannaud91lics}, a powerful generalization of
G{\"o}del's higher-order primitive recursion of higher types. Following
the path initiated by Breazu-Tannen and Gallier on the one hand, and
Okada on the other hand, termination of calculi based on the general
schema was proved by using computability arguments as
well~\cite{jouannaud91lics,jouannaud97tcs,barbanera94lics}.  The
general schema was then reformulated by Blanqui, Jouannaud and
Okada~\cite{blanqui99rta,blanqui02tcs} in order to incorporate
computability arguments directly in its definition, opening the way to
new generalizations. G{\"o}del's system $T$ can be generalized in two
ways, by introducing type constructors and dependent types, yielding
the Calculus of Constructions, and by introducing strictly positive
inductive types.  Both together yield the Calculus of Inductive
Constructions~\cite{paulin93tlca}, the theory underlying the Coq
system~\cite{coq8.0}, in which rewrite rules like strong
elimination operate on types, raising new difficulties. Blanqui gave a
generalization of the general schema which includes the Calculus of
Inductive Constructions as a particular case under the name of
Calculus of Algebraic Constructions~\cite{blanqui05mscs,blanqui05fi}.

The general schema, however, is too simple to analyze complex calculi
defined by higher-order rewrite rules such as encodings of logics. For
that purpose, Jouannaud and Rubio generalized the recursive path
ordering to the higher-order case, yielding the higher-order recursive
path ordering (HORPO)~\cite{jouannaud99lics}.  The RPO
well-foundedness proof follows from Kruskal's tree theorem, but no
such theorem exists in presence of a binding construct, and it is not
at all clear that such a theorem may exist. What is remarkable is
that computability arguments fit with RPO's recursive structure. When
applied to RPO, these arguments result in a new, simple,
well-foundedness proof of RPO. One could even argue that this is the
\emph{first} well-foundedness proof of RPO, since Dershowitz showed
\emph{more}: well-orderedness.

Combining the general schema and the HORPO is indeed easy because
their termination properties are both based on computability
arguments. The resulting relation, HORPO with closure, combines an
ordering relation with a membership predicate. In this paper, we
reformulate and improve a recent idea of Blanqui~\cite{blanqui06horpo}
by defining a new version of the HORPO with closure which integrates
smoothly the idea of the general schema into HORPO in the form of a
new ordering definition.

So far, we have considered the kind of higher-order rewriting defined by
using first-order pattern matching as in the calculus of
constructions. These orderings need to contain $\beta$- and
$\eta$-reductions.  Showing termination of higher-order rewrite rules
based on higher-order pattern matching, that is, rewriting modulo
$\beta$ and $\eta$ now used as equalities, turns out to require simple
modifications of HORPO~\cite{jouannaud06rta}. We will therefore concentrate here on
higher-order orderings containing $\beta$- and $\eta$-reductions.

We introduce higher-order algebras in Section~\ref{s:preliminaries}.
In Section~\ref{s:computability}, we recall the computability argument
for this variation of the simply typed lambda calculus. Using a 
computability argument again, we show in Section~\ref{s:rpo} that RPO
is well-founded.  We introduce the general schema in
section~\ref{s:closure}, and the HORPO in Section~\ref{s:horpo} before
to combine both in Section~\ref{s:newhorpo}. We end up with related
work and open problems in the last two sections.

\section{Higher-Order Algebras}
\label{s:preliminaries}

The notion of a higher-order algebra given here is the monomorphic
version of the notion of polymorphic higher-order algebra defined
in~\cite{jouannaud06jacm}. Polymorphism has been ruled out for simplicity.

\subsection{Types, Signatures and Terms}
\label{types}
Given a set $\Sort$ of \emph{sort symbols} of a fixed arity, denoted by
$s:*^n\Dra *$, the set $\Type$ of
\emph{types} is generated from these sets by the arrow
constructor:
\[ 
\begin{array}{c}
\Type := s(\Type^n) ~|~ (\Type \ra \Type) \\
\mbox{for } s: *^n\Dra *~\in\Sort
\end{array}
\] 
Types headed by $\ra$ are \emph{arrow types} while the others
are \emph{basic types}.  \emph{Type declarations} are expressions
of the form $\sigma_1 \times \cdots \times \sigma_n \ra \sigma$, where
$n$ is the \emph{arity} of the type declaration, and $\sigma_1,
\ldots,\sigma_n, \sigma$ are types.  A type declaration is {\em
first-order} if it uses only sorts, otherwise \emph{higher-order}.

We assume given a set of function symbols which are meant to be algebraic
operators. Each function symbol $f$ is equipped with a type
declaration $f\colon\sigma_1 \times\cdots \times \sigma_n \ra
\sigma$. We use $\F_n$ for the set of function symbols of arity $n$.
$\F$ is a \emph{first-order signature} if all its type
declarations are first-order, and a higher-order signature otherwise.

The set of \emph{raw terms} is generated from
the signature ${\cal F}$ and a denumerable set $\X$ of
variables according to the grammar: 
\[ 
\Term := 
   \X \mid (\lambda \X . \Term) \mid @(\Term,\Term) \mid \F(\Term,\ldots,\Term).
\]
Terms generated by the first two grammar rules are called
\emph{algebraic}.  Terms of the form $\lambda x.u$ are called
\emph{abstractions} while terms of the form $@(u,v)$ are called
\emph{applications}. The term $@(u,\vect{v})$ is called a (partial)
\emph{left-flattening} of $@(\dots @(@(u,v_1),v_2),\dots, v_n)$,
with
$u$ being possibly an application itself.  Terms other than
abstractions are said to be \emph{neutral}.  We denote by $\Var{t}$
($\BVar{t}$) the set of free (bound) variables of $t$.  We may assume
for convenience (and without further notice) that bound variables in a
term are all different, and are different from the free ones.

Terms are identified with finite labeled trees by considering $\lambda
x .$, for each variable $x$, as a unary function symbol.  {\em
Positions} are strings of positive integers, the empty string $\rootp$
denoting the root position.  The {\it subterm} of $t$ at position $p$
is denoted by $t|_p$, and by $t[u]_p$ the result of replacing $t|_p$
at position $p$ in $t$ by $u$. We write $s\gtsubt u$ if $u$ is a
strict subterm of $s$.  We use $t[~]_p$ for a term with a hole, called
a context.  The notation $\vect{s}$ will be ambiguously used to denote
a list, a multiset, or a set of terms $s_1,\ldots,s_n$.

\subsection{Typing Rules}
\label{tr}
Typing rules restrict the set of terms by constraining them to follow
a precise discipline. Environments are sets of pairs written
$x:\sigma$, where $x$ is a variable and $\sigma$ is a type. Let
$\Dom{\Gamma}=\{x \mid x:\sigma\in\Gamma \mbox { for some type }
\sigma\}$.  We assume there is a unique pair of the form $x:\sigma$
for every variable $x\in\Dom{\Gamma}$.  Our typing judgments are
written as $\Gamma \turnstyle M : \sigma$ if the term $M$ can be
proved to have the type $\sigma$ in the environment $\Gamma$.  A term
$M$ has type $\sigma$ in the environment $\Gamma$ if $\Gamma
\turnstyle M : \sigma$ is provable in the inference system of
Figure~\ref{f:typing}. A term $M$ is typable in the environment
$\Gamma$ if there exists a type
$\sigma$ such that $M$ has type $\sigma$ in the environment
$\Gamma$. A term $M$ is typable if it is typable in some environment
$\Gamma$.  Note that function symbols are uncurried, hence must come
along with all their arguments.

\begin{figure}
\begin{center}
\fbox{
$
\begin{array}{ccc}
\begin{array}{c}
\newinfRULE
{Variables}      
{x : \sigma \in \Gamma}
{\Gamma \turnstyle x : \sigma}
\end{array}
&\quad&
\begin{array}{c}
\newinfRULE
{Functions}
{\begin{array}{c}
f : \sigma_1 \times \ldots \times \sigma_n \ra \sigma\\
\Gamma \turnstyle t_1 : \sigma_1 ~\ldots~  \Gamma \turnstyle t_n : \sigma_n
\end{array}}
{\Gamma \turnstyle f(t_1,\ldots,t_n): \sigma}
\end{array}
\\\\
\begin{array}{c}
\newinfRULE
{Abstraction}
{\Gamma \cup \{x : \sigma \} \turnstyle t : \tau}
{\Gamma \turnstyle (\lambda x:\sigma.t) : \sigma \ra \tau}
\end{array}
&&
\begin{array}{c}
\newinfRULE
{Application}
{\Gamma \cup \{x : \sigma \} \turnstyle s : \sigma \ra \tau~~~~
\Gamma \turnstyle t : \sigma }
{\Gamma \turnstyle @(s, t) : \tau}
\end{array}
\\\\
\end{array}
$
}
\end{center}
\caption{Typing judgments in higher-order algebras}\label{f:typing}
\end{figure}

\subsection{Higher-Order Rewrite Rules}
Substitutions are written as in $\{x_1:\sigma_1 \mapsto (\Gamma_1,
t_1),\ldots, x_n:\sigma_n \mapsto (\Gamma_n,t_n)\}$ where, for every
$i\in[1..n]$, $t_i$ is assumed different from $x_i$ and
$\Gamma_i\turnstyle t_i:\sigma_i$. We also assume that
$\bigcup_i{\Gamma_i}$ is an environment.
We often write $x\mapsto t$ instead of
$x:\sigma\mapsto (\Gamma,t)$, in particular when $t$ is ground.  We
use the letter $\gamma$ for substitutions and postfix notation for
their application.  Substitutions behave as endomorphisms defined on
free variables.

\noindent%
A (possibly higher-order) {\it term rewriting system} is a set of
rewrite rules $R=\{\Gamma_i \turnstyle l_i \ra r_i:\sigma_i\}_i$, 
where $l_i$ and $r_i$ are higher-order terms such that $l_i$ and $r_i$
have the same type $\sigma_i$ in the environment $\Gamma_i$.  Given a
term rewriting 
system $R$, a term $s$ rewrites to a term $t$ at position
$p$ with the rule $l\ra r$ and the substitution $\gamma$, written
$\displaystyle s \lrps{p}{l\ra r} t$, or simply $s \ra_R t$, if $s|_p
= l\gamma$ and $t = s[r\gamma]_p$.

A term $s$ such that $\displaystyle s\lrps{p}{R} t$ is called {\em
$R$-reducible}.  The subterm $s|_p$ is a \emph{redex} in $s$, and $t$
is the \emph{reduct} of $s$.  Irreducible terms are said to be in {\it
$R$-normal form}. A substitution $\gamma$ is in $R$-normal form if
$x\gamma$ is in $R$-normal form for all $x$.  We denote by
$\displaystyle \lrps{*}{R}$ the reflexive, transitive closure of the
rewrite relation $\displaystyle \lrps{}{R}$.

Given a rewrite relation $\lrps{}{}$, a term $s$ is strongly
normalizing if there is no infinite sequence of rewrites issuing from
$s$.  The rewrite relation itself is {\it strongly normalizing}, or
{\it terminating}, if all terms are strongly normalizing, in which case
it is called a \emph{reduction}.

Three particular higher-order equation schemas originate from the $\lambda$-calculus,
$\alpha$-, $\beta$- and $\eta$-equality:
\begin{center}
$
\begin{array}{rcll}
\lambda x . v  & =_\alpha & \lambda y . v\{x\mapsto y\}
 & \mbox{if}~ y\not\in\BVar{v}~\cup~ (\Var{v}\setminus\{x\})\\
@(\lambda x . v, u) & \lrps{}{\beta} & v\{x \mapsto u\} &\\
\lambda x . @(u, x) & \lrps{}{\eta} & u & \mbox{if}~ x\not\in\Var{u}
\end{array}
$
\end{center}

As usual, we do not distinguish $\alpha$-convertible terms.
$\beta$- and $\eta$-equalities are used as reductions, which is
indicated by the long-arrow symbol instead of the equality symbol.
The above rule-schemas define a rewrite system which is known to be
terminating, a result proved in Section~\ref{s:computability}.

\subsection{Higher-Order Reduction Orderings}
\label{ro}
We will make intensive use of well-founded orderings, using the
vocabulary of rewrite systems for orderings, for proving strong
normalization properties. For our purpose, an \emph{ordering}, usually
denoted by $\geq$, is a reflexive, symmetric, transitive relation
compatible with $\alpha$-conversion, that is, $s=_\alpha t\geq
u=_\alpha v$ implies $s\geq v$, whose strict part $>$ is itself
compatible. We will essentially use strict orderings, and hence, the
word ordering for them too. We will also make use of order-preserving
operations on relations, namely multiset and lexicographic extensions,
see~\cite{dershowitz88}.

\emph{Rewrite orderings} are \emph{monotonic} and \emph{stable}
orderings, \emph{reduction orderings} are in addition {\em
well-founded}, while \emph{higher-order reduction orderings} must also
contain $\beta$- and $\eta$-reductions. Monotonicity of $>$ is defined
as $u > v$ implies $s[u]_p > s[v]_p$ for all contexts
$s[~]_p$. Stability of $>$ is defined as $u > v$ implies $s\gamma >
t\gamma$ for all substitutions $\gamma$. Higher-order reduction
orderings are used to prove termination of rewrite systems including
$\beta$- and $\eta$-reductions by simply comparing the left hand and
right hand sides of the remaining rules.

\section{Computability}
\label{s:computability}

Simply minded arguments do not work for showing the strong
normalization property of the simply typed lambda-calculus, for
$\beta$-reduction increases the size of terms, which precludes
an induction on their size, and preserves their types, which seems to
preclude an induction on types.

Tait's idea is to generalize the strong normalization property in
order to enable an induction on types. To each type $\sigma$, we
associate a subset $\cand{\sigma}$ of the set of terms, called the
\emph{computability predicate} of type $\sigma$, or set of
\emph{computable terms} of type $\sigma$.  Whether $\cand{\sigma}$
contains only typable terms of type $\sigma$ is not really important,
although it can help intuition.  What is essential are the properties
that the family of predicates should satisfy:

\begin{compactenum}[(i)]
\item computable terms are strongly normalizing;
\item reducts of computable terms are computable;
\item a neutral term $u$ is computable iff all its reducts are
  computable;
\item $u:\sigma\ra\tau$ is computable iff so is $@(u,v)$ for all
  computable $v$.
\end{compactenum}

A (non-trivial) consequence of all these properties can be added to
smooth the proof of the coming Main Lemma:

\begin{compactenum}[(i)]\setcounter{enumi}{4}
\item \label{fifth}
$\lambda x.u$ is computable iff so is $u\{x\mapsto
v\}$ for all computable $v$.
\end{compactenum}

Apart from (\ref{fifth}), the above properties refer to
$\beta$-reduction via the notions of \emph{reduct} and \emph{strong
  normalization} only. Indeed, various computability predicates found
in the literature use the same definition parametrized by the
considered reduction relation.

There are several ways to define a computability predicate by taking
as its definition some of the properties that it should satisfy. For
example, a simple definition by induction on types is this:

$s:\sigma\in\cand{\sigma}$ for $\sigma$ basic iff $s$ is strongly normalizing;

$s:\theta\ra\tau\in\cand{\sigma\ra\tau}$ iff
$@(s,u):\tau\in\cand{\tau}$ for every $u:\theta\in\cand{\theta}$.

\noindent
An alternative for the case of basic type is:
\begin{displaymath}
  s:\sigma\in\cand{\sigma} \mbox{~iff~} \forall t:\tau~.~s\lrps{}{}t
  \mbox{~then~}
  t\in\cand{\tau} .
\end{displaymath}
This formulation defines the predicate as a
fixpoint of a monotonic functional. Once the predicate is defined, it
becomes necessary to show the computability properties. This uses an induction on
types in the first case or an induction on the definition of the
predicate in the fixpoint case.

Tait's strong normalization proof is based on the following key lemma:

\begin{lemma}[Main Lemma]
Let $s$ be an arbitrary term and $\gamma$ be an arbitrary computable
substitution. Then $s\gamma$ is computable.
\end{lemma}

\begin{proof}
By induction on the structure of terms.
\begin{enumerate}
\item
$s$ is a variable: $s\gamma$ is computable by assumption on $\gamma$.
\item
$s=@(u,v)$. Since $u\gamma$ and $v\gamma$ are computable by induction
hypothesis, $s\gamma=@(u\gamma,v\gamma)$ is computable by
computability property (iv).
\item
$s=\lambda x . u$.  By computability property (v), $s\gamma=\lambda
x.u\gamma$ is computable iff $u\gamma\{x\mapsto v\}$ is computable for
an arbitrary computable $v$.  Let now $\gamma'=\gamma\cup\{x\mapsto
v\}$. By definition of substitutions for abstractions,
$u\gamma\{x\mapsto v\}=u\gamma'$, which is usually ensured by
$\alpha$-conversion.  By assumptions on $\gamma$ and $v$, $\gamma'$ is
computable, and $u\gamma'$ is therefore computable by the main induction
hypothesis.
\qed
\end{enumerate}
\end{proof}

Since an arbitrary term $s$ can be seen as its own instance by the
identity substitution, which is computable by computability property
(iii), all terms are computable by the Main Lemma, hence strongly
normalizing by computability property (i).

\section{The Recursive Path Ordering and Computability}
\label{s:rpo}

In this section, we restrict ourselves to first-order algebraic terms.
Assuming that the set of function symbols is equipped with an ordering
relation $\geF$, called \emph{precedence}, and a status function $stat$,
writing $stat_f$ for $stat(f)$, we recall the definition of the
recursive path ordering:

\begin{definition}
\label{d:rpo}
$s\gtrpo t$  iff
\begin{enumerate}
\item
\label{rposubt}
$s=f(\vect{s})$ with $f\in\F$,
and 
$ u \displaystyle{\gerpo} t$ for some
$u\in\vect{s}$

\item
\label{rpoprec}
$s=f(\vect{s})$ with $f\in\F$, and
$t=g(\vect{t})$ with $f\gtF g$, and $A$

\item
\label{rpomul}
$s=f(\vect{s})$ and $t=g(\vect{t})$ with $f=_\F g\in\Mul$, and
$\vect{s}~(\displaystyle{\gtrpo})_{mul}~ \vect{t}$

\item
\label{rpolex}
$s=f(\vect{s})$ and $t=g(\vect{t})$ with $f=_\F g\in\Lex$, and
$\vect{s}~(\displaystyle{\gtrpo})_{lex}~ \vect{t}$ and $A$
\end{enumerate}
$$\mbox{where } A=\forall v\in\vect{t}. ~ s\gtrpo v \quad\mbox{ and }\quad
              s\gerpo t \mbox{ iff } s\gtrpo t \mbox{ or } s=t$$
\end{definition}

We now show the well-foundedness of $\gtrpo$ by using Tait's
method. Computability is defined here as strong normalization, 
implying computability property (i). We prove the computability
property:

(vii) Let $f\in\F_n$ and $\vect{s}$ be computable terms.
Then $f(\vect{s})$ is computable.

\begin{proof}
The restriction of $\succ_{rpo}$ to terms smaller than or equal to the
terms in $\vect{s}$ w.r.t. $\succ_{rpo}$ is a well-founded ordering
which we use for building an outer induction on the pairs
$(f,\vect{s})$ ordered by $(>_\F, (\succ_{rpo})_{stat_f})_{lex}$. This
ordering is well-founded, since it is built from well-founded
orderings by using mappings that preserve well-founded orderings.
  
We now show that $f(\vect{s})$ is computable by proving that $t$ is
computable for all $t$ such that $f(\vect{s}) \succ_{rpo} t$. This
property is itself proved by an (inner) induction on $|t|$, and by
case analysis upon the proof that $f(\vect{s}) \succ_{rpo} t$.
\begin{enumerate}
\item
subterm case: $\exists u\in\vect{s}$ such that $u \succ_{rpo} t$. By assumption,
$u$ is computable, hence so is its reduct $t$.
\item
precedence case: $t=g(\vect{t})$, $f>_\F g$, and $\forall v\in\vect{t}$,
$s\succ_{rpo} v$. By inner induction, $v$ is computable, hence
so is $\vect{t}$. By outer induction, $g(\vect{t})=t$ is computable.
\item
multiset case: $t=f(\vect{t})$ with $f\in Mul$, and
$\vect{s}(\succ_{rpo})_{mul} \vect{t}$. By definition of the multiset
extension, $\forall v\in\vect{t},~\exists u\in\vect{s}$ such that
$u\succeq_{rpo} v$. Since $\vect{s}$ is a vector of computable terms
by assumption, so is $\vect{t}$. We conclude by outer induction that
$f(\vect{t})=t$ is computable.
\item
lexicographic case: $t=f(\vect{t})$ with $f\in Lex$,
$\vect{s}(\succ_{rpo})_{lex} \vect{t}$, and $\forall
v\in\vect{t},~s\succ_{rpo} v$. By inner induction, $\vect{t}$ is strongly
normalizing, and by outer induction, so is $f(\vect{t})=t$.
\qed
\end{enumerate}
\end{proof}

\noindent
The well-foundedness of $\gtrpo$ follows from computability property (vii).

\section{The General Schema and Computability}
\label{s:closure}

As in the previous section, we assume that the set of function symbols
is equipped with a precedence relation $\geF$ and a status function  $stat$.

\begin{definition} 
\label{d:closure} 
The \emph{computability closure} $\CS(t=f(\vect{t}))$, with $f\!\in\!\F$,
is the set $\CS(t,\emptyset)$, s.t. $\CS(t,{\cal V})$, with ${\cal
V}\cap\Var{t}=\emptyset$, is the smallest set of typable terms
containing all variables in ${\cal V}$ and terms in $\vect{t}$, 
closed under:
\begin{enumerate} 
\item
\label{csubterm}
subterm of basic type: let $s\in\CS(t,{\cal V})$, and $u$ be a subterm
 of $s$ of basic type $\sigma$ such that $\Var{u}\subseteq \Var{t}$;
 then $u\in\CS(t,{\cal V})$;
\item
\label{cprecedence}
precedence: let $f\gtF g$, and
$\vect{s}\in\CS(t,{\cal V})$; then $g(\vect{s})\in\CS(t,{\cal V})$;
\item
\label{crecursive}
recursive call: let $f(\vect{s})$ be a term such that terms in $\vect{s}$
belong to $\CS(t,{\cal V})$ and $\vect{t}
(\lrps{}{\beta}\cup\gtsubt)_{stat_f} \vect{s}$; then
$g(\vect{s})\in\CS(t,{\cal V})$ for every $g =_\F f$;
\item
\label{capplication}
application: let $s :
\sigma_1\ra\ldots\ra\sigma_n\ra\sigma\in\CS(t,{\cal V})$ and
$u_i:\sigma_i \in\CS(t,{\cal V})$ for every $i\in[1..n]$; then
$@(s,u_1,\ldots,u_n)\in\CS(t,{\cal V})$;
\item
\label{cabstraction}
abstraction: let $s\in\CS(t,{\cal V}\cup\{x\})$ for some
$x\notin\Var{t}\cup{\cal V}$;
then $\lambda x . s \in\CS(t,{\cal V})$;
\item 
\label{creduction}
reduction: let $u \in \CS(t,{\cal V})$, and
$u\lrps{}{\beta\cup\gtsubt} v$; then $v\in \CS(t,{\cal V})$.
\end{enumerate}
\end{definition}

We say that a rewrite system $R$ satisfies the \emph{general schema} iff
$$\mbox{$r\in\CS(f(\vect{l}))$ for all $f(\vect{l})\ra r\in R$}
$$

We now consider computability with respect to the rewrite relation
$\lrps{}{R}\cup\lrps{}{\beta}$, and add the computability property (vii)
whose proof can be easily adapted from the previous one.
We can then add a new case in Tait's Main Lemma, for terms headed by
an algebraic function symbol. As a consequence, the relation
$\lrps{}{\beta}\cup\lrps{}{R}$ is strongly normalizing.

\begin{example}[System T]
\label{recursor}
We show the strong normalization of G{\"o}del's system $T$ by showing that
its rules satisfy the general schema. This is clear for the first rule
by the base Case of the definition. For the second rule, we have:
$V\in\CS(rec(s(X),U,V))$ by base Case; $s(X)\in\CS(rec(s(X),U,V))$ by
base Case again, and 
$X\in\CS(rec(s(X),U,V))$ by Case~\ref{cprecedence}, assuming $rec\gtF s$; $U\in\CS(rec(s(X),U,V))$
by base Case, hence all arguments of the recursive call are in
$\CS(rec(s(X),U,V))$.  Since $s(X)\gtsubt X$ holds, we have
$rec(X,U,V)\in\CS(rec(s(X),U,V))$. Therefore, we conclude with
\mbox{$@(V,X,rec(X,U,V))\in\CS(rec(s(X),U,V))$} by Case~\ref{capplication}.
\end{example}

\section{The Higher-Order Recursive Path Ordering}
\label{s:horpo}

\subsection{The  Ingredients}
\label{ssto}

\begin{itemize}
\item
A quasi-ordering on types $\geS$ called \emph{the type ordering}
satisfying the following properties:
\begin{enumerate}
\item
\label{wf}
\emph{Well-foundedness}: $\gtS$ is well-founded;

\item
\label{poa}
\emph{Arrow preservation}:
$\tau\ra\sigma \eqS \alpha~\mbox{iff}~
\alpha=\tau'\ra\sigma',~\tau'\eqS \tau~\mbox{and}~\sigma \eqS \sigma';$

\item
\label{sa}
\emph{Arrow decreasingness}:
$\tau\ra\sigma \gtS \alpha ~\mbox{implies}~ \sigma \geS \alpha \mbox{ or }
\alpha=\tau'\ra\sigma', \tau'\eqS \tau$ \mbox{ and $\sigma \gtS \sigma'$;}

\item
\label{ma}
\emph{Arrow monotonicity}:
$\tau\geS\sigma~\mbox{implies}~ \alpha\ra\tau\geS\alpha\ra\sigma  ~\mbox{and}~
\tau\ra\alpha\geS\sigma\ra\alpha;$
\end{enumerate}
A convenient type ordering is obtained by restricting the subterm
property for the arrow in the RPO definition.

\item
A quasi-ordering $\geF$ on ${\cal F}$, called the \emph{precedence},
  such that $\gtF$ is well-founded.

\item
A \emph{status} $stat_f\in\{Mul,Lex\}$ for every symbol $f\in\F$.
\end{itemize}

The higher-order recursive path ordering (HORPO) operates on typing
judgments. To ease the reading, we will however forget the
environment and type unless necessary.  Let
\[A=\forall v \in\vect{t}~~
s \displaystyle{\gthorpo} v ~\mbox{or}~ 
u \displaystyle{\gthorpo} v
~\mbox{for some}~u\in\vect{s}\] 

\begin{definition}
\label{horpodef}
Given two judgments $\Gamma \judgF s: \sigma$ and $\Sigma \judgF t: \tau$,
\[
s \gthorpo t ~\mbox{iff}~ \sigma \geS \tau ~\mbox{and}
\]
\begin{enumerate}
\item
\label{horposubterm}
$s=f(\vect{s})$ with $f\in\F$,
and 
$u \displaystyle{\gehorpo} t$ for some
$u\in\vect{s}$

\item
\label{horpoprec}
$s=f(\vect{s})$ with $f\in\F$, and
$t=g(\vect{t})$ with $f\gtF g$, and $A$

\item
\label{horpomul}
$s=f(\vect{s})$ and $t=g(\vect{t})$ with $f=_\F g\in\Mul$, and 
$\vect{s}~(\displaystyle{\gthorpo})_{mul}~\vect{t}$

\item
\label{horpolex}
$s=f(\vect{s})$ and $t=g(\vect{t})$ with $f=_\F g\in\Lex$, and
$\vect{s}~(\displaystyle{\gthorpo})_{lex}~\vect{t}$ and $A$

\item
\label{horposubtapp}
$s=@(s_1,s_2)$, and $s_1 \displaystyle{\gehorpo} t$ 
or $s_2 \displaystyle{\gehorpo} t$

\item
\label{horposubtlam}
$s=\lambdax:\alpha.u$ with $x\not\in\Var{t}$, and 
$u \displaystyle{\gehorpo} t$

\item
\label{horpoprecapp}
$s=f(\vect{s})$ with $f\in\F$,
$t=@(\vect{t})$ is a partial left-flattening of $t$, and $A$

\item
\label{horpopreclam}
$s=f(\vect{s})$ with $f\in\F$,
$t=\lambdax:\alpha.v$ with $x\not\in\Var{v}$ and 
$s\displaystyle{\gthorpo} v$

\item
\label{horpoapp}
$s=@(s_1,s_2)$, $t=@(\vect{t})$,  and $\{s_1, s_2\} 
(\displaystyle{\gthorpo})_{mul}~ \vect{t}$

\item
\label{horpolambda}
$s=\lambdax:\alpha.u,~t=\lambdax:\beta.v$, $\alpha\eqS\beta$, and
$u \displaystyle{\gthorpo} v$ 

\item
\label{horpobeta}
$s=@(\lambdax:\alpha.u,v)$ and 
$u\{x\mapsto v\} \displaystyle{\gehorpo} t$

\item
\label{horpoeta}
\label{number}
$s=\lambdax:\alpha. @(u,x), ~x\not\in\Var{u}$ and 
$u \displaystyle{\gehorpo} t$
\end{enumerate}
\end{definition}

\begin{example}[System T]
The new proof of strong normalization of System $T$ is even simpler.
For the first rule, we apply Case~\ref{horposubterm}. For the second,
we apply Case~\ref{horpoprecapp}, and show recursively
that $rec(s(X),U,V)\gthorpo V$ by Case~\ref{horposubterm},
$rec(s(X),U,V)\gthorpo X$ by Case~\ref{horposubterm} applied twice,
and $rec(s(X),U,V)\gthorpo rec(X,U,V)$ by Case~\ref{horpomul},
assuming a multiset status for $rec$, which follows from the comparison
$s(X)\gthorpo X$ by Case~\ref{horposubterm}.
\end{example}

The strong normalization proof of HORPO is in the same style as the
previous strong normalization proofs, although technically more
complex~\cite{jouannaud06jacm}. This proof shows that HORPO and the
general schema can be combined by replacing the membership
$u\in\vect{s}$ used in case~\ref{horposubterm} by the more general
membership $u\in\CS(f(\vect{s}))$. It follows that the HORPO mechanism
is inherently more expressive than the closure mechanism.

Because of Cases~\ref{horpobeta} and~\ref{horpoeta}, HORPO is not
transitive. Indeed, there are examples for which the proof of
$s\gthorpo^+ t$ requires guessing a middle term $u$ such that
$s\gthorpo u$ and $u\gthorpo t$. Guessing a middle term when necessary
is automated in the implementations of HORPO and HORPO with closure
available from the web page of the first two authors.

\section{Unifying HORPO and the Computability Closure}
\label{s:newhorpo}

A major advantage of HORPO over the general schema is its recursive
structure. In contrast, the membership to the computability closure is
undecidable due to its Case~\ref{crecursive}, but does not involve any
type comparison. To combine the advantages of both, we now incorporate
the closure construction into the HORPO as an ordering.  Besides, we
also incorporate the property that arguments of a type constructor are
computable when the \emph{positivity condition} is satisfied as it is
the case for inductive types in the Calculus of Inductive
Constructions~\cite{paulin93tlca,blanqui05fi}.

\[
\begin{array}{lcl}
\begin{array}{@{}c@{}}

{\displaystyle s: \sigma \gthorpo t: \tau}\quad\mbox{iff}\\
  \Var{t}\subseteq\Var{s}\quad\mbox{and}\\\\

\begin{array}{@{}ll@{}}
1.&
s=f(\vect{s}) \mbox{ and } {\displaystyle s\gtchorpoX{\emptyset} t}\\

2.&
s=f(\vect{s}) \mbox{ and } \sigma\geS\tau \mbox{ and}\\

&\begin{array}{@{}ll@{}}

(a)&
t=g(\vect{t}),~ f\gtF g \mbox{ and } A\\

(b)&
t=g(\vect{t}),~ f=_\F g,\\
& {\displaystyle \vect{s}(\gthorpo)_{stat_f}\vect{t}} \mbox{ and } A\\

(c)&
t=@(t_1,t_2) \mbox{ and } A
\end{array}\\

3.&
s=@(s_1,s_2), \sigma\geS\tau \mbox{ and}\\

&\begin{array}{@{}ll@{}}
(a)&
t=@(t_1,t_2) \mbox{ and}\\
& {\displaystyle \{s_1,s_2\} (\gthorpo)_{mul} \{t_1,t_2\}}\\

(b)&
{\displaystyle s_1\gehorpo t}$ or ${\displaystyle s_2\gehorpo t}\\

(c)&
s_1=\lambda x.u \mbox{ and}\\
&{\displaystyle u\{x\mapsto s_2\}\gehorpo t}
\end{array}\\

4.&
s=\lambdax:\alpha.u, \sigma\geS\tau \mbox{ and}\\
&\begin{array}{ll}

(a)&
t=\lambdax:\beta.v, \alpha\eqS\beta\\
& \mbox{and } {\displaystyle u\gthorpo v}\\

(b)&
x\not\in\Var{t}$ and ${\displaystyle u\gehorpo t}\\

(c)&
u=@(v,x), x\not\in\Var{v}\\
& \mbox{and } {\displaystyle v\gehorpo t}
\end{array}
\end{array}\\\\

\mbox{where } A=\forall v \in\vect{t}:\\
s \displaystyle{\gthorpo} v \mbox{ or }
\exists u\in\vect{s}:~u \displaystyle{\gthorpo} v
\end{array}

&\quad&

\begin{array}{@{}c@{}}

{\displaystyle s=f(\vect{s})\gtchorpoX{\vect{X}} t}\quad\mbox{iff}\\\\

\begin{array}{@{}ll@{}}
1.&
t\in\vect{X}\\

2.&
\exists s_i\in Acc(s):~ s_i \gechorpoX{\vect{X}} t\\

3.&
t=g(\vect{t}),~ f\gtS g \mbox{ and}\\
& \forall v\in\vect{t}:~
{\displaystyle s\gtchorpoX{\vect{X}}v}\\

4.&
t=g(\vect{t}),~ f\eqS g,\\
& \forall v\in\vect{t}~: 
{\displaystyle s\gtchorpoX{\vect{X}}v} \mbox{ and}\\
& {\displaystyle Acc(s) (\gthorpo)_{stat_f} \lambda\vect{X}.\vect{t}}\\

5.&
t=@(u,v),\\
& {\displaystyle s\gtchorpoX{\vect{X}} u} \mbox{ and }
{\displaystyle s\gtchorpoX{\vect{X}} v}\\

6.&
t=\lambdax:\alpha.u \mbox{ and}\\
& {\displaystyle s\gtchorpoX{\vect{X}\cdot\{x:\alpha\}} u}
\end{array}\\\\

\mbox{where } s_i \in Acc(f(\vect{s}))\\
(s_i \mbox{ is accessible in } s)\\
\mbox{ iff}\\

\begin{array}{@{}ll@{}}
1.&
s \mbox{ is the left hand side of}\\
& \mbox{an ancestor goal } s\gthorpo u\\

2.&
s \mbox{ is the left hand side of the}\\
& \mbox{current goal } s\gtchorpoX{} u, \mbox{ and,}\\
& \mbox{either} f:\vect{\sigma}\ra\sigma \mbox{ and}\\
& \sigma \mbox{ occurs only positively in } \sigma_i.
\end{array}
\end{array}
\end{array}
\]

\begin{example}
\label{ex:brouwer}
We consider now the type of Brouwer's ordinals defined from the type
$\Nat$ by the equation $Ord=0 \uplus s(Ord) \uplus lim(\Nat\ra
Ord)$. Note that $Ord$ occurs positively in the type $\Nat\ra Ord$,
and that $\Nat$ must be smaller or equal to $Ord$.
The recursor for the type $Ord$ is defined as:

\begin{eqnarray*}
  rec(0,U,V,W) &\ra& U\\
  rec(s(X),U,V,W) &\ra& @(V,X,rec(X,U,V,W))\\
  rec(lim(F),U,V,W) &\ra& @(W,F,\lambda n.rec(@(F,n),U,V,W))
\end{eqnarray*}

\noindent
We skip the first two rules and concentrate on the third:
\begin{compactenum}
  \item \label{goal:1}
    $rec(lim(F),U,V,W) \gthorpo @(W,F,\lambda n.rec(@(F,n),U,V,W))$\\
    which, by Case~1 of $\gthorpo$ is replaced by the new goal:
  \item \label{goal:2}
    $rec(lim(F),U,V,W) \gtchorpoX{\emptyset} @(W,F,\lambda
    n.rec(@(F,n),U,V,W))$\\
    By Case~5 of $\gtchorpoX{}$, these three goals become:
  \item \label{goal:3}
    $rec(lim(F),U,V,W) \gtchorpoX{\emptyset} W$
  \item \label{goal:4}
    $rec(lim(F),U,V,W) \gtchorpoX{\emptyset} F$
  \item \label{goal:5} $rec(lim(F),U,V,W) \gtchorpoX{\emptyset}
    \lambda
    n.rec(@(F,n),U,V,W)$\\
    Since $rec(lim(F),U,V,W)$ originates from Goal~\ref{goal:1},\\
    Goal~\ref{goal:3} disappears by Case~2, while
    Goal~\ref{goal:4} becomes:
  \item \label{goal:6}
    $lim(F)\gtchorpoX{\emptyset} F$\\
    which disappears by the same Case since $F$ is accessible in
    $lim(F)$.\\
    thanks to the positivity condition. By Case~6, Goal~\ref{goal:5}
    becomes:
  \item \label{goal:7}
    $rec(lim(F),U,V,W) \gtchorpoX{\{n\}} rec(@(F,n),U,V,W)$\\
    Case~4 applies with a lexicographic status for $rec$, yielding 5 goals:
  \item \label{goal:8}
    $rec(lim(F),U,V,W) \gtchorpoX{\{n\}} @(F,n)$
  \item \label{goal:9}
    $rec(lim(F),U,V,W) \gtchorpoX{\{n\}} U$
  \item \label{goal:10}
    $rec(lim(F),U,V,W) \gtchorpoX{\{n\}} V$
  \item \label{goal:11}
    $rec(lim(F),U,V,W) \gtchorpoX{\{n\}} W$
  \item \label{goal:12} $\{lim(F),U,V,W\} (\gthorpo)_{lex}
    \{\lambda n.@(F,n), \lambda n.U, \lambda n.V, \lambda n.W\}$\\
    Goals~\ref{goal:9}, \ref{goal:10}, \ref{goal:11} disappear by
    Case~2, while, by Case~5\\
    Goal~\ref{goal:8} generates (a variation of) the solved
    Goal~\ref{goal:4} and the new sub-goal:
  \item \label{goal:13}
    $rec(lim(F),U,V,W) \gtchorpoX{\{n\}} n$\\
    which disappears by Case~1. We are left with Goal~\ref{goal:12},
    which reduces to:
  \item \label{goal:14}
    $lim(F) \gthorpo \lambda n.@(F,n)$\\
    which, by Case~1 of $\gthorpo$, then~6 and~5 of $\gtchorpoX{}$
    yields successively:
  \item \label{goal:15}
    $lim(F) \gtchorpoX{\emptyset} \lambda n.@(F,n)$
  \item \label{goal:16}
    $lim(F) \gtchorpoX{\{n\}} @(F,n)$\\
    which, by Case~5, generates (a variation of) the Goal~\ref{goal:6} and the last goal:
  \item \label{goal:17}
    $lim(F) \gtchorpoX{\{n\}} n$\\
    which succeeds by Case~1, ending the computation.
  \end{compactenum}
\end{example}

To show the strong normalization property of this new definition of
$\horpo$, we need a more sophisticated predicate combining the
predicates used for showing the strong normalization of
HORPO~\cite{jouannaud06jacm} and CAC~\cite{blanqui05mscs}.  We have not
done any proof yet, but we believe that it is well-founded.

It is worth noting that the ordering $\horpo$ defined here is in one
way less powerful than the one defined in Section~\ref{s:horpo} using
the closure definition of Section~\ref{s:closure} because it does not
accumulate computable terms for later use anymore.  Instead, it
deconstructs its left hand side as usual with rpo, and remembers very
few computable terms: the accessible ones only. On the other hand, it
is more powerful since the recursive case~4 of the closure uses now
the full power of $\horpo$ for its last comparison instead of simply
$\beta$-reduction (see~\cite{jouannaud06jacm}). Besides, there is no
more type comparison in Case~1 of the definition of $\horpo$, a key
improvement which remains to be justified formally.

\section{Related Work}
\label{s:related}

Termination of higher-order calculi has recently attracted quite a lot
of attention. The area is building up, and mostly, although not
entirely, based on reducibility techniques.

The case of conditional rewriting has been recently investigated by
Blanqui~\cite{blanqui06lpar}. His results are presented in this
conference.

Giesl's dependency pairs method has been generalized to higher-order
calculi by using reducibility techniques as described
here~\cite{sakai06,blanqui06hodp}. The potential of this line of work
is probably important, but more work in this direction is needed to
support this claim.

Giesl~\cite{giesl06rta} has achieved impressive progress for the case of
combinator based calculi, such as Haskell programs, by transforming
all definitions into a first-order framework, and then proving
termination by using first-order tools. Such transformations do not
accept explicit binding constructs, and therefore, do not apply to
rich $\lambda$-calculi such as those considered here. On the other hand,
the relationship of these results with computability deserves investigation.

An original, interesting work is Jones's analysis of the flux of
redexes in pure lambda-calculus~\cite{jones04rta}, and its use for
proving termination properties of functional programs. Whether this
method can yield a direct proof of finite developments in pure
$\lambda$-calculus should be investigated.  We also believe that his
method can be incorporated to the HORPO by using an interpretation on
terms instead of a type comparison, as mentioned in Conclusion.

Byron Cook, Andreas Podelski and Andrey Ribalchenko~\cite{podelski}
have developed a quite different and impressive method based on
abstract interpretations to show termination of large imperative
programs. Their claim is that large programs are more likely to be
shown terminating by approximating them before to make an
analysis. Note that the use of a well-founded ordering can be seen as
a particular analysis. Although impressive, this work is indeed quite
far from our objectives.

\section{Conclusion}
\label{s:conclusion}

We give here a list of open problems which we consider important. We
are ourselves working on some of these.  The higher-order recursive
path ordering should be seen as a firm step to undergo further
developments in different directions, some of which are listed below.
\begin{itemize}
\item Two of them have been investigated in the first order framework:
  the case of associative commutative operators, and the use of
  interpretations as a sort of elaborated precedence operating on
  function symbols. The first extension has been carried out for the
  general schema~\cite{blanqui03rta}, and the second for a weak form
  of HORPO~\cite{rubio01}. Both should have an important impact for
  applications, hence deserve immediate attention.
\item Enriching the type system with dependent types, a problem
  considered by Wa\-{\l}u\-kie\-wicz~\cite{walukiewicz00lfm} for the
  original version of HORPO in which types were compared by a
  congruence. Replacing the congruence by HORPO recursively called on
  types as done in~\cite{jouannaud06jacm} for a simpler type
  discipline raises technical difficulties.  The ultimate goal here is
  to generalize the most recent versions of the ordering including the
  present one, for applications to the Calculus of Inductive
  Constructions.
\item HORPO does not contain and is not a well-order for the subterm
  relationship. However, its definition shows that it satisfies a weak
  subterm property, namely property $A$. It would be theoretically
  interesting to investigate whether some Kruskal-like theorem holds
  for higher-order terms with respect to the weak subterm property.
  This could yield an alternative, more abstract way of hiding away
  computability arguments.
\end{itemize}

\end{document}

%% file: macros.tex
\newcommand{\lambdax}{\lambda \hspace{-0.30mm} x}

\newcommand{\Sort}{{\cal S}}

\newcommand{\Type}{{\cal T}_{\cal S}}
\newcommand{\gttype}{>_{\Type}}
\newcommand{\getype}{\geq_{\Type}}

\newcommand{\eqtype}{=_{\Type}}
\newcommand{\gtS}{\gttype}
\newcommand{\geS}{\getype}

\newcommand{\eqS}{\eqtype}

\newcommand{\Term}{\TT}

\newcommand{\gtF}{>_{\cal F}}

\newcommand{\geF}{\geq_{\cal F}}

\newcommand{\cand}[1]{[\![ #1 ]\!]}

\newcommand{\vect}[1]{\overline{#1}}

\newcommand{\turnstyle}{~\vdash~}

\newcommand{\ra}{\rightarrow}

\newcommand{\Dra}{\Rightarrow}

\newcommand{\Nat}{\mathbb{N}}

\newcommand{\TT}{{\cal T}}
\newcommand{\F}{{\cal F}}

\newcommand{\X}{{\cal X}}

\newcommand{\Var}[1]{{\cal V}ar({#1})}
\newcommand{\BVar}[1]{{\cal BV}ar({#1})}

\newcommand{\rootp}{\mbox {\footnotesize $\Lambda$}}

\newcommand{\Dom}[1]{{\cal D}om({#1})}

\newcommand{\lrps}[2]{\mathop{\longrightarrow}^{#1}_{#2}}

\newcommand{\gtpo}{\succ}
\newcommand{\gepo}{\succeq}

\newcommand{\gtsubt}{\rhd}

\newcommand{\horpo}{\mathop{\gtpo}_{horpo}}
\newcommand{\gthorpo}{\mathop{\gtpo}_{horpo}}
\newcommand{\gehorpo}{\mathop{\gepo}_{horpo}}

\newenvironment{ruleset}
               {$$ \begin{array}{llcr}}{\end{array} $$ }

\newcommand{\Farc}[2]{\displaystyle \frac{#1}{#2}}

\newcommand{\newinfRULE}[3]
{\textbf{#1:}\\
\Farc{#2}{#3}}

\newcommand{\Mul}{Mul}
\newcommand{\Lex}{Lex}

\newcommand{\CS}{{\cal CC}}

\newcommand{\gtrpo}{\mathop{\gtpo}_{rpo}}
\newcommand{\gerpo}{\mathop{\gepo}_{rpo}}
\newcommand{\gtchorpoX}[1]{\mathop{\gtpo}^{#1}_{comp}}
\newcommand{\gechorpoX}[1]{\mathop{\gepo}^{#1}_{comp}}

\newcommand{\judgF}{\turnstyle_{\!\!\!\!\!\Sigma}~}